\documentclass[aps,prl,twocolumn,superscriptaddress, amsmath]{revtex4-1}
\usepackage{amssymb,graphicx,xspace,color}
\newcommand{\steffen}[1]{\textcolor{black}{#1}}

\begin{document}

\title{Anomalous Nernst effect and field-induced Lifshitz transition in Weyl semimetals TaP and TaAs} 
\author{F. Caglieris}
\affiliation{Leibniz Institute for Solid State and Materials Research, 01069 Dresden, Germany}
\author{C. Wuttke}
\affiliation{Leibniz Institute for Solid State and Materials Research, 01069 Dresden, Germany}
\author{S. Sykora}
\affiliation{Leibniz Institute for Solid State and Materials Research, 01069 Dresden, Germany}
\author{V. S\"uss}
\affiliation{Max Planck Institute for Chemical Physics of Solids, 01187 Dresden, Germany}
\author{C. Shekhar}
\affiliation{Max Planck Institute for Chemical Physics of Solids, 01187 Dresden, Germany}
\author{C. Felser}
\affiliation{Max Planck Institute for Chemical Physics of Solids, 01187 Dresden, Germany}

\author{B. B\"uchner}
\affiliation{Leibniz Institute for Solid State and Materials Research, 01069 Dresden, Germany}
\affiliation{Institut f\"{u}r Festk\"{o}rperphysik, TU Dresden, 01069 Dresden, Germany}
\affiliation{Center for Transport and Devices, TU Dresden, 01069 Dresden, Germany}
\author{C. Hess}
\affiliation{Leibniz Institute for Solid State and Materials Research, 01069 Dresden, Germany}
\affiliation{Center for Transport and Devices, TU Dresden, 01069 Dresden, Germany}

\date{\today}
\begin{abstract}

The discovery of Weyl fermions in transition metal monoarsenides/phosphides without inversion symmetry represents an exceptional breakthrough in modern condensed matter physics. However, exploring the inherent nature of these quasiparticles is experimentally elusive because most of the experimental probes rely on analysing Fermi arc topology or controversial signatures such as the appearance of the chiral anomaly and the giant magnetoresistance. Here we show that the prototypical type-I Weyl semimetals TaP and TaAs possess a giant anomalous Nernst signal with a characteristic saturation plateau beyond a critical field which can be understood as a direct consequence of the finite Berry curvature originating from the Weyl points. Our results thus promote the Nernst coefficient as an ideal bulk probe for detecting and exploring the fingerprints of emergent Weyl physics.

\end{abstract}

\maketitle

\begin{figure}
\includegraphics[width=\columnwidth]{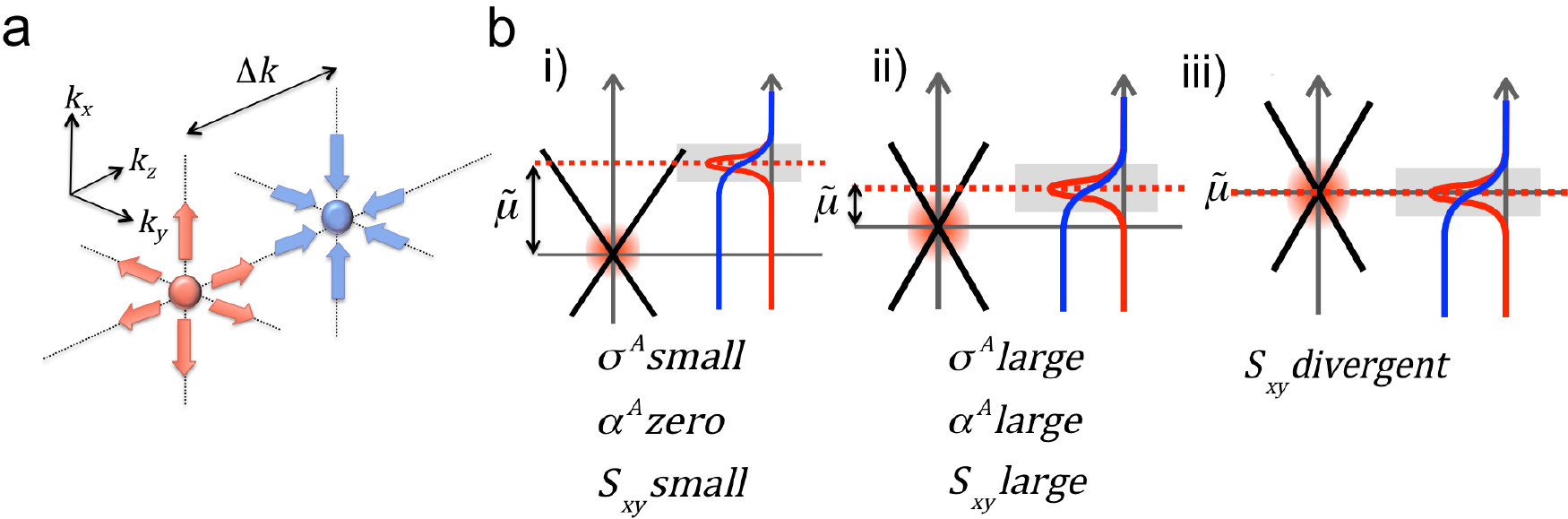}

\caption{\label{chempot}a) Schematic picture of the Berry curvature originated from right-handed (blue) and left-handed (red) Weyl points. The Weyl points always appear as a pair with opposite chirality separated by $\Delta k$ in the $k$-space and they act as source or sink of Berry curvature (outward or inward arrows). b) \steffen{Schematic picture of the folding between the energy dependent Fermi distribution function (blue) and entropy density (red) with the diverging Berry curvature (red shaded area) near the Weyl point. The grey shaded area marks the thermal energy $k_BT$. The ANE is particularly large and sensitive to a variation of the chemical potential if the grey and red shaded areas overlap, i.~e.~$\tilde{\mu} \approx k_BT$, which is highlighted in the cases ii) and iii).}}
\end{figure}

Three-dimensional (3D) topological Weyl semimetals (TWSs) \cite{Yang:2015aa, Xue1501092, PhysRevX.5.011029, Huang:2015aa, PhysRevX.5.031013, Xu613,Xu:2015aa} are characterized by a peculiar electronic structure at half way between a 3D analogue of graphene and topological insulators. Indeed, TWSs present bulk Weyl fermions, chiral particles that disperse linearly along all three momentum directions across the corresponding Weyl points. The Weyl points appear always in pairs, separated in momentum space as consequence of spin-orbit coupling and breaking of the time-reversal symmetry or inversion symmetry (IS). Hence, the chiral Weyl fermions experience a finite Berry curvature for which the Weyl points act as source or sink \cite{PhysRevX.5.011029} (Fig.~\ref{chempot}a). Recently, intrinsic Weyl states have been observed in transition metal monoarsenides/phosphides NbP, NbAs, TaP, TaAs, with naturally broken IS \cite{Yang:2015aa,Xue1501092, PhysRevX.5.011029, Huang:2015aa, PhysRevX.5.031013, Xu613,Xu:2015aa}.

The experimental evidence for the presence of Weyl states essentially concerns the investigation of the topological surface state, i. e. the so-called Fermi arcs. More indirect signatures emerge from magneto-transport measurements, which have recently revealed extremely high carrier mobility \cite{arxivMob,PhysRevB.93.121112,Shekhar:2015aa} and negative longitudinal magnetoresistance \cite{PhysRevX.5.031023,Li:2016aa,Zhang:2016aa,Arnold:2016aa} in different candidate materials. These have been indeed interpreted as fingerprints of particles with defined chirality, but unfortunately leave room for ambiguity. Since a finite Berry curvature at the Fermi level acts on fermions analogous to a magnetic field, the presence of sizeable so-called anomalous transverse transport quantities, viz.~the anomalous Hall effect (AHE) and the anomalous Nernst effect (ANE) has been predicted as a more direct, unambiguous proof of a finite Berry curvature and thus the existence of Weyl nodes close to the Fermi level \cite{PhysRevB.93.035116, PhysRevB.96.195119, PhysRevB.90.165115,PhysRevLett.93.226601,PhysRevLett.97.026603}. Investigating the Nernst effect is particularly interesting for revealing these sought-after anomalous contributions because in ordinary metals its normal contribution vanishes \cite{Wang2001,Behnia2009,Sondheimer1936}, in contrast to the Hall effect, where the normal contribution is always finite and scales reciprocally with the carrier density.
In this paper we show that prototypical Weyl semimetals TaP and TaAs indeed possess a giant anomalous component in the Nernst coefficient $S_{xy}$ with a characteristic saturation plateau beyond a critical magnetic field $B_s$, which univocally certifies these materials as Weyl semimetals in the bulk. Using a minimal dispersion model of Weyl fermions we explain this unique field dependence as a consequence of a field-induced shift of the chemical potential, resulting in a Lifshitz transition of the particular Fermi surface sheets which enclose the relevant Weyl points. Taking specific material parameters from band structure calculations we estimate $B_s$ in consistency with our experimental results. Our work thus points out the Nernst effect as a smoking-gun experiment for the identification of the emerging Weyl physics in all the candidate materials \cite{Xu:2015aa,PhysRevB.94.121113,Deng:2016aa,PhysRevB.94.165145,PhysRevB.95.241108}.

\begin{figure}
\includegraphics[width=\columnwidth]{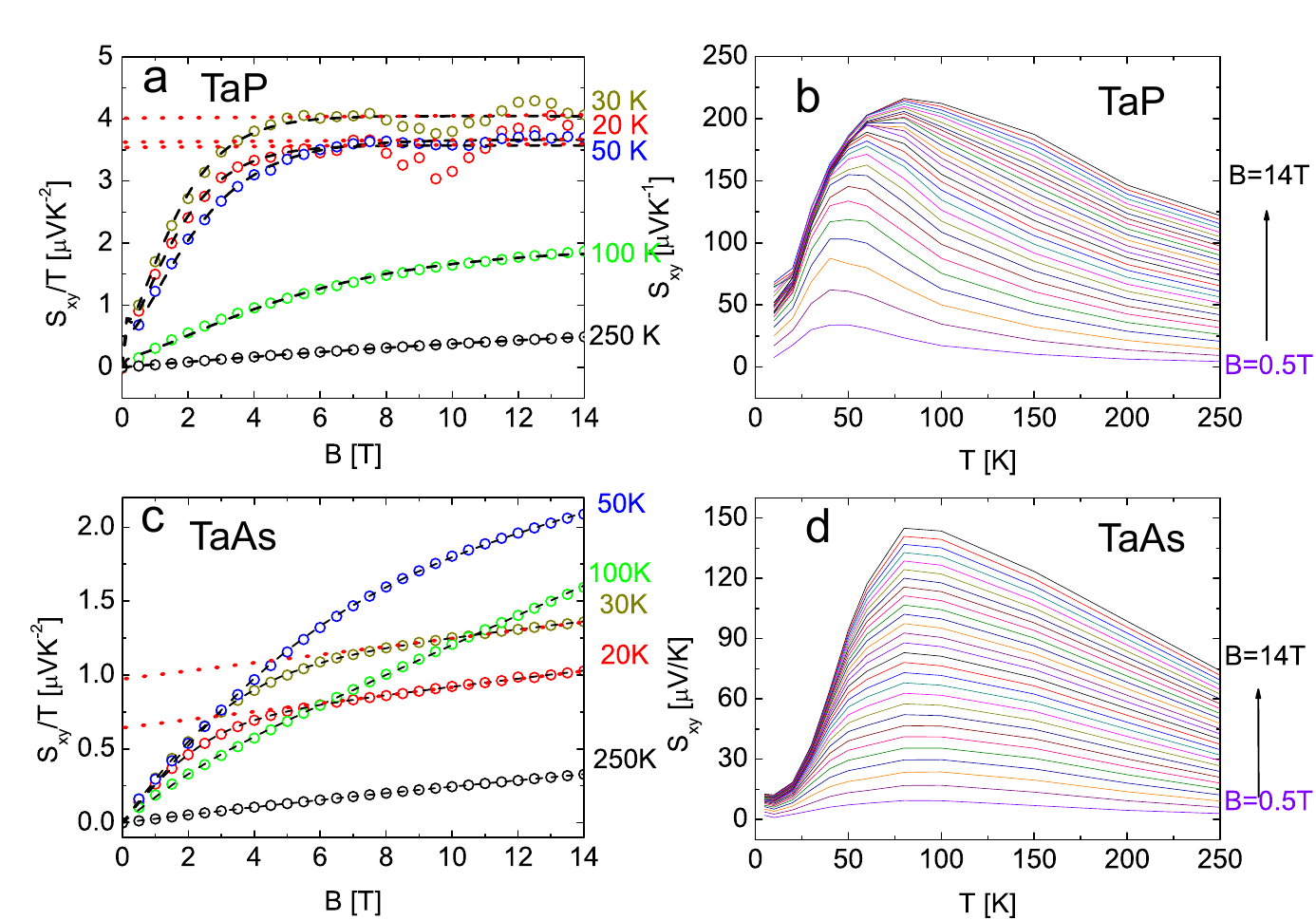}

\caption{\label{data} Nernst effect data for TaP (panels a and b) and TaAs (panels c and d). Panels a and c show the Nernst coefficient normalized to the temperature $S_{xy}/T$ as a function of $B$. In both of the compounds an anomalous contribution appears as a tendency to saturation. Panels b and d show the Nernst coefficient $S_{xy}$  as a function of the temperature $T$. In both of the compounds we find a crossover between two different temperature regimes which appears as a change of slope with a broadened maximum. The red dotted lines in panels a and c represent guides for the eyes. The black dashed lines in panels a and c show the results of the phenomenological fit, which well reproduces the experimental data for all the $T$. Shubnikov-de Haas oscillations appear in the $S_{xy}/T$ vs $B$ curves of TaP at low temperature and they are visible up to 60 K \cite{SM}.}
\end{figure}

Fig.~\ref{data}a shows the experimental data \cite{SM} of the magnetic field ($B$) dependence of the Nernst coefficient normalized to the temperature, $S_{xy}/T$, for a single crystal of TaP, in the field range $B=0-14$ T at selected temperatures. First of all, the absolute value of $S_{xy}/T$ is large for all the temperatures, reaching the magnitude of what is typically addressed as 'giant' Nernst effect. This represents a strong violation of the so-called Sondheimer's cancellation, which causes the suppression of the Nernst signal in standard single-band materials \cite{0034-4885-79-4-046502}. Remarkably, for $T=20$~K, 30 K and 50 K, after an initial increase, $S_{xy}/T$ develops an extended plateau beyond a saturation field $B_s \approx 4$ T, superimposed with quantum oscillations. The establishment of flat plateaus represents a strong departure from the conventional theory of transport, which predicts the normal Nernst coefficient $S^N_{xy}$ evolving directly proportional to the magnetic field $B$ or to its inverse $B^{-1}$ in the low and high field limit, respectively \cite{Liang:2013aa,0034-4885-79-4-046502}, as reproduced by Eq. 2. The saturating behavior of $S_{xy}/T$ therefore implies the presence of an anomalous, i.e., a magnetic field independent component in the Nernst signal. It resembles the behavior of ferromagnetic solids \cite{PhysRevLett.93.226601, PhysRevLett.101.117208, PhysRevB.90.054422}, but is strongly unusual in non-magnetic materials with the exception of the Dirac semimetal Cd$_3$As$_2$, where similar plateaus have been observed\cite{PhysRevLett.118.136601}.

The inspection of the temperature evolution of the data reveals that such anomalous component dominates the Nernst effect even up to 50~K. However, for $T=100$~K, even if the tendency of $S_{xy}/T$ to saturate is still persistent, a constant value is not reached (at least for $B<$14 T) and for $T=250$~K the curve is almost completely dominated by a linear component, explainable in terms of a normal contribution according to the semiclassical theory. Interestingly, the crossover between a low-$T$ and a high-$T$ regime, dominated by an anomalous and a normal contribution respectively, is reproduced in the temperature dependence of $S_{xy}$ for different fields (Fig.~\ref{data}b). In fact, in these curves $S_{xy}$ shows a broadened maximum at around 80 - 100  K for high fields which shifts to lower temperature by decreasing $B$. We mention that a similar maximum exists also in the analogous compound NbP \cite{PhysRevB.97.161404}. However, the characteristic plateaus, which identify the ANE reported here for TaP, do not clearly appear in that case \cite{PhysRevB.97.161404}.

In analogy to TaP, we performed Nernst measurements on a sample of TaAs. Fig.~\ref{data}c shows the $B$-dependence of $S_{xy}/T$. Unlike the case of TaP, for $T=20$~K, 30 K and 50 K, the $S_{xy}/T$ vs $B$ curves do not reach a constant value. Even at these low temperatures a $B$-linear component is apparently overimposed to a large saturating anomalous part, leading to a linear drift of $S_{xy}/T$ instead of a plateau for high fields. Remarkably, this normal contribution becomes already dominant at around 100 K and persists up to 250 K, where the anomalous part is completely unobservable. The difference in the normal component of the Nernst coefficient in the two compounds is not surprising if we consider the strong sensitivity of the conventional Nernst effect even to subtleties of a multi-band electronic structure \cite{0034-4885-79-4-046502}.

We extract the qualitative $T$-dependence of the amplitude of the anomalous part of the Nernst coefficient by using a phenomenological model, which decomposes in a normal $S^N_{xy}$ and an anomalous $S^A_{xy}$ contributions \cite{PhysRevLett.118.136601}:
\begin{eqnarray}
{S_{xy}} &=& S^N_{xy}+S^A_{xy}, \label{S_xy} \\
{S^N_{xy}} &=& S^N_0 \frac{\mu_e B}{1+(\mu_e B)^2} , \label{S_xyN} \\
{S^A_{xy}} &=& S^A_0 \mathrm{tanh} (B/B_s),
\label{S_xyA} 
\end{eqnarray}
with $\mu_e$ the average mobility,  $S_0^N$ and $S_0^A$ the amplitudes of the normal and the anomalous part, respectively, and $B_s$ the saturation field \cite{PhysRevLett.118.136601}. The normal contribution as given by Eq.~\eqref{S_xyN} can be derived from standard theory of transport \cite{Liang:2013aa}. We also note that multiband effects can enhance the normal Nernst contribution as a consequence of ambipolar transport but these effects also behave linearly as a function of magnetic field. The field-dependence of the anomalous part can be modeled by the empirical formula~\eqref{S_xyA} which is widely accepted to reproduce the saturation effect at high fields \cite{PhysRevLett.118.136601}.

\begin{figure}

\includegraphics[width=\columnwidth]{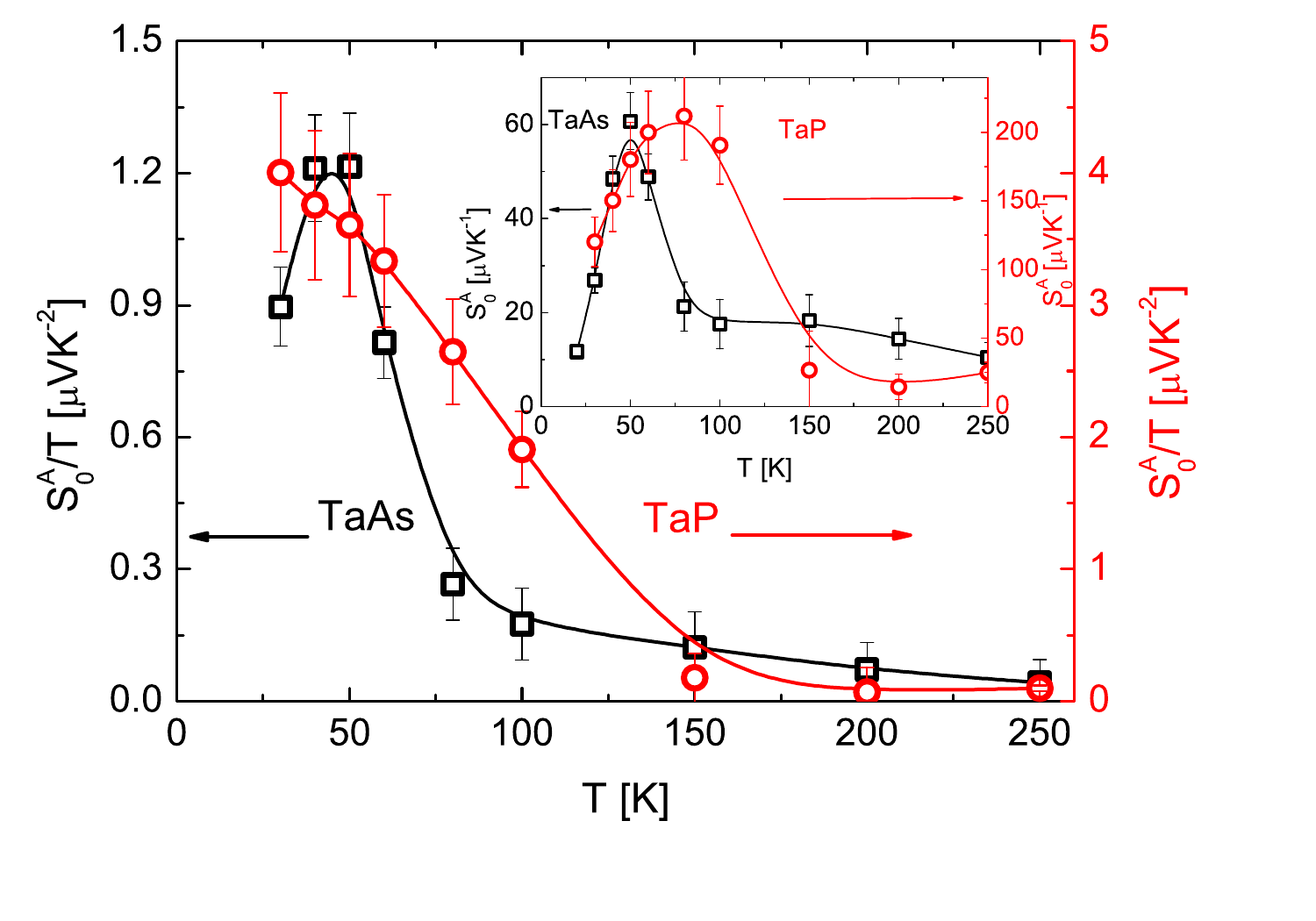}
\caption{\label{Fig_S_anomalous} Temperature dependence of the amplitude of the anomalous part of the Nernst coefficient normalized to the temperature, $S^A_0/T$, for TaP (red) and TaAs (black). A crossover between two different regimes is clearly observable in the abrupt change of $S^A_0/T$ that occurs in TaP and TaAs around 150 K and 100 K, respectively. Inset: Temperature dependence of $S^A_0$ for TaP (red) and TaAs (black) without temperature normalisation. A broadened maximum exists at around $T=80$ K and $T=50$ K in TaP and TaAs, respectively.}
\end{figure}

The black dashed lines in Fig.~\ref{data}a and c show the results of the fit of $S_{xy}/T$ for the sample of TaP and TaAs, respectively. The phenomenological model works well for all $T$, allowing the isolation of the anomalous component $S^A_0$. Fig.~\ref{Fig_S_anomalous} shows the $T$-dependence of $S^A_0/T$ for TaP and TaAs. In both the compounds, a rapid decrease of around one order of magnitude indicates the crossover from a low-$T$ to a high-$T$ regime where the anomalous component changes from strong to weak, respectively. The decrease starts at around 150 K for TaP and at around 100 K for TaAs. It is noteworthy that the transition occurs at the same temperature regime that characterizes a strong change of the Hall resistivity $\rho_{xy}$ which in both compounds results in a very high average Hall mobility at low $T$ \cite{arxivMob,PhysRevB.93.121112,Shekhar:2015aa, Hu:2016aa}. This high mobility has been attributed to the suppression of back-scattering due to the emergence of Weyl fermions \cite{arxivMob,PhysRevB.93.121112,Shekhar:2015aa}.

After having established the presence and $T$-dependence of the ANE in TaP and TaAs we turn now to \steffen{ an explanation of the characteristic field independent plateau in the ANE.} 
The Nernst coefficient $S_{xy} = E_y / (-\nabla_x T)$ can be generally expressed in terms of the thermoelectric tensor $\alpha_{ij}$ and charge conductivity tensor $\sigma_{ij}$ in the form,
\begin{eqnarray*}
S_{xy} &=& \frac{\alpha_{xy} \sigma_{xx} - \alpha_{xx} \sigma_{xy}}{\sigma_{xx}^2 + \sigma_{xy}^2}.
\end{eqnarray*}
It is well established within the Boltzmann approach that anomalous terms proportional to the integrated Berry curvature projected to the direction of the magnetic field ($\Omega_n^{z}$ with $n$ a band index) generally arise in the transverse coefficients  \cite{PhysRevB.93.035116}:

\begin{eqnarray}
\label{sigma}
\sigma_{xy}^A &=&  \frac{e^2}{\hbar} \sum_n \int \frac{d^3k}{(2\pi)^3} \Omega_n^{z}({\bf k}) f_{n,{\bf k}} \\
\alpha_{xy}^A &=&  \frac{k_B e}{\hbar} \sum_n \int \frac{d^3k}{(2\pi)^3} \Omega_n^{z}({\bf k}) s_{n,{\bf k}}.
\label{alpha}
\end{eqnarray}
Here, $f_{n,{\bf k}} = f(E_{n{\bf k}})$ is the Fermi distribution function and $s_{n,{\bf k}} = - f_{n,{\bf k}} \ln f_{n,{\bf k}} - \left[ (1 - f_{n,{\bf k}}) \ln (1 - f_{n,{\bf k}}) \right]$ the entropy density for the dispersion $E_{n,{\bf k}}$ of the conduction electron band $n$.

Closer analysis of Eqs. (4,5) shows that the ANE due to a finite $\Omega_n^{z}$ is large if the energy difference between the chemical potential $\mu$ and the Weyl point is of the same order of magnitude or is smaller than the thermal energy $k_B T$ (see cases ii) and iii) in Fig.~\ref{chempot}b) \cite{SM}. This is due to the folding of the momentum dependence between the diverging Berry curvature $\Omega_n^{z}({\bf k})$ and $f_{n,{\bf k}}$ (or $s_{n,{\bf k}}$) which becomes particularly strong when $\mu$ moves towards the Weyl point \cite{Note2}.
The ANE therefore becomes sensitive to both, variations of the chemical potential $\mu$ and of the temperature $T$. More specifically, it is expected to grow upon reducing $\mu$ as well upon increasing $T$ from $k_BT\ll{\tilde{\mu}}$ toward $k_BT\approx{\tilde{\mu}}$, where ${{\tilde{\mu}}}$ measures the distance between the Weyl node and $\mu$.

\begin{figure}
\includegraphics[width=\columnwidth]{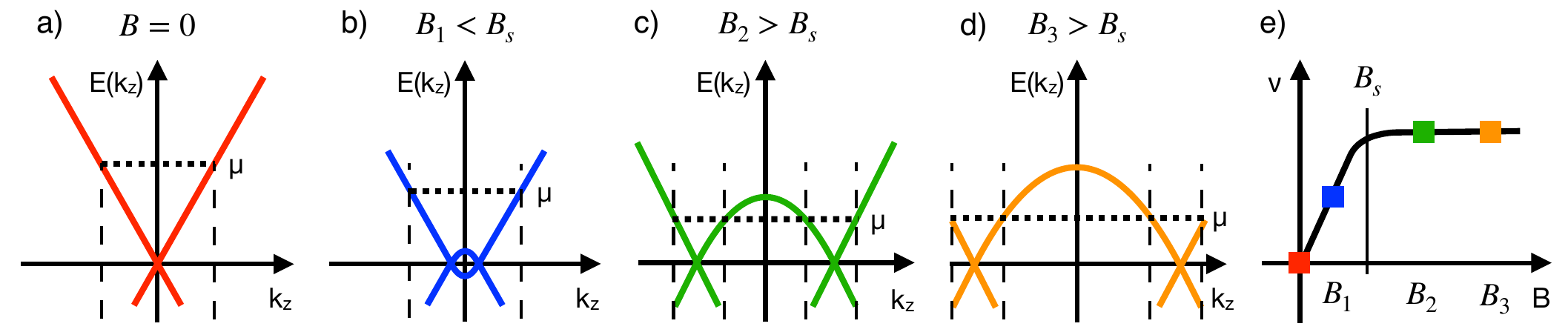}
\caption{
\label{Fig_theory}
Schematic behavior of the chemical potential $\mu$ in a Dirac/Weyl semimetal. a) In zero-field the energy dispersion shows only one single Dirac node which is placed in the $\Gamma$-point. b) Application of magnetic field $B$ leads to the appearance of two nodes whose $k$-space separation is proportional to $B$. A shift of $\mu$, indicated by the dotted lines, is induced to conserve the particle number under variation of $B$. The downshift of $\mu$  is responsible for the observed strong  variation of the Nernst coefficient for $B < B_s$, where $B_s$ is the saturation field of the Nernst coefficient. c) For $B > B_s$ a Lifshitz transition occurs and two separated Fermi surfaces appear. d) By further rising $B > B_s$ the Weyl cones distance increases without changing the $k$-space volume. Hence, the conservation of the particle number (area between the dashed lines) is already achieved if $\mu$ remains almost fixed. As a consequence, the anomalous Nernst coefficient $S_{xy}^A$ will also not change as highlighted in panel e). 
% \steffen{{\it Lower right panel}: Schematic picture of the folding between the energy dependent Fermi distribution function (blue) and entropy density (red) with the diverging Berry curvature (red shaded area) near the Weyl point. The grey shaded area marks the thermal energy $k_BT$. The ANE is particularly large and sensitive to a variation of the chemical potential if the grey and red shaded areas overlap, i.~e.~$\tilde{\mu} \approx k_BT$.}
}
\end{figure}

\paragraph*{Magnetic field-induced shift of the chemical potential}
Based on a minimal dispersion model for Weyl fermions in a magnetic field we now argue that the position of the chemical potential is closely related to the $B$-value and also to the topology of the Fermi surface. %This relation leads immediately to an explanation of the observed plateaus and is the main purpose of the following theoretical considerations.
If a magnetic field is applied to a Dirac or Weyl semimetal with the Dirac or Weyl points in the vicinity of the chemical potential, the latter typically experiences a $B$-induced shift. This can be inferred from Fig.~\ref{Fig_theory} where for simplicity the effect of a magnetic field applied on a doubly degenerate Dirac cone along the $z$-direction is considered.
\steffen{Fig.~\ref{Fig_theory}a} shows the $k_z$-dispersion around a Dirac point for $B=0$. The chemical potential $\mu$ takes a certain position (black dotted horizontal line) above the node, determined by the number of conducting electrons, which corresponds at zero-temperature to the number of the occupied states below the Fermi level.
The application of a magnetic field along the $z$-direction splits the Dirac cone into two Weyl cones which immediatly leads to a finite ANE (\steffen{Fig.~\ref{Fig_theory}b}). The magnitude of this ANE remains small as long as $k_BT\ll\tilde{\mu}$ (Fig.~\ref{chempot}b i) \cite{SM}.
Increasing the magnetic field, however, leads to a further separation of the Weyl nodes where the inevitable downshift of unoccupied states causes a reduction of the chemical potential $\mu$ in order to fulfill the condition of a fixed particle number. Thus, the distance between the Weyl points and the chemical potential, $\tilde{\mu}$, is reduced resulting, as discussed above, in an enhancement of the ANE, in agreement with the experiment. For larger fields, $\mu$ is pushed to the region where the Weyl cones are separated, such that the Fermi surface topology has changed (\steffen{Fig.~\ref{Fig_theory}c}), leading to the appearance of two distinct Fermi surfaces. In contrast to the behavior at low fields, in this regime the chemical potential remains nearly constant since a further increasing magnetic field simply leads to further separation of the Weyl nodes but not to a change of the dispersion near the Weyl nodes as sketched in \steffen{Fig.~\ref{Fig_theory}d}. As we will discuss in detail below, this mechanism works also if the starting situation at B=0 is not a degenerate Dirac cone but two preformed Weyl nodes, as in case of inversion symmetry breaking TWS, since the driving parameter is the additional shift in momentum space induced by the field and the consequent variation of the $k$-space volume.

This behavior qualitatively explains the observed saturation of the Nernst coefficient when the magnetic field is larger than some material characteristic value $B_s$ as schematically presented in \steffen{Fig.~\ref{Fig_theory}e}. To compare this scenario with our actual experimental results we estimate the value of $B_s$. To this end, we have numerically evaluated $\mu$ as a function of $B$ based on the minimal dispersion model $E_{\bf k}=\pm \sqrt{v_F^2 (k_x^2 + k_y^2) + (gB \pm v_F k_z)^2}$ \cite{PhysRevB.93.035116}. Specifically, we have calculated the Fermi volume $\Delta V$ as a function of $\mu$ and $B$ using the integral $\Delta V = \int d^3k f_{\bf k} (\mu,B)$. Then we have solved the obtained equation numerically for $\mu$ leading to a functional relation between $\mu$ and $B$ from which we have read off the saturation field, $B_s$, defined by  constant $\mu$ for $B > B_s$. From this approach we find the following relation between  $B_s$, the total Fermi volume $\Delta V$, and the Fermi velocity $v_F$:
\begin{eqnarray}
\label{B_s}
B_s &=& v_F \frac{m_e}{e}  \root 3 \of {\frac{3\Delta V}{2\pi}}0.745,
\end{eqnarray}
where $m_e$ denotes the electron mass and $e$ the elementary charge. The number 0.745 in Eq.~(\ref{B_s}) corresponds to the ratio of the saturated value of the chemical potential compared to its original value, \steffen{$\mu_s / \tilde{\mu} = 0.745$, where both $\mu_s$ and $\tilde{\mu}$ are related to the energy of the Weyl point.} 
For  Weyl semimetals such as TaP and TaAs, $\mu_s / \tilde{\mu}$ assumes a material specific different value of similar magnitude.
From band structure calculations \cite{Lee2015} one finds for these materials preformed Weyl points along $k_x$ (with a relative distance $\Delta k_x$ of the order of 10$^{-2}$\AA$^{-1}$ in the $k$-space).
The \steffen{anomalous $S_{xy}$ measured here} has been observed with the magnetic field applied in the $z$-direction, inducing a further separation of the Weyl nodes along $k_z$, where for both TaAs and TaP $\Delta k_z = 2gB/v_F$ ($g$ is the Lande factor) is of the order of $4 \cdot 10^{-3}$\AA$^{-1}$ at $B=10$~T. \steffen{Band structure calculations show that in both materials there are two relevant pairs of Weyl points, usually denoted as $W_1$ and $W_2$, which are close to the Fermi level \cite{Lee2015}. To obtain our saturation effect at least one pair needs to be enclosed by a non-separated Fermi surface. In TaP, while $W_2$ has a separated zero-field Fermi surface, for $W_1$ the saddle point between the pair of Weyl points is roughly 0.015 eV below the Fermi level ($\tilde{\mu} \approx -0.05$ eV) indicating a non-separated Fermi surface for $B=0$. This property allows to obtain the discussed saturation effect through a shift of the chemical potential at moderate field values. In the specific situation of TaP the above numbers suggest  a somewhat enhanced $\mu_s/\mu=0.751$ and thus, following Eq.~\ref{B_s}, $B_s=3.3$ T}, where  $\Delta V \approx 10^{-9}$ \AA$^{-3}$ and $v_F \approx 10^{5}$~m/s have been used as approximate values \cite{Lee2015}. For TaAs, on the other hand, for both Weyl points the Fermi surface is already separated according to band structure calculations. Nevertheless, the described saturation effect is observed also in this material. We attribute \steffen{this materials specific discrepancy between the experiment and the quantitative predictions} to  shortcomings of the precision of band structure calculations. \steffen{Note that the saddle point energy of $W_1$ in TaAs is calculated to a value of approximately 0.01 eV which is slightly above the Fermi level. However, a precise description of one-particle states at such small energy leading to the typical small Fermi volumes in TWSs requires an exceptional high $k$ point sampling in the band structure calculation. Therefore, we believe that an improvement in such a direction could lead to a correction of the saddle point energy leading to a situation similar to TaP.}

\paragraph*{Impact of temperature}
Due to the broadening of $f_{n,{\bf k}}$ (or $s_{n,{\bf k}}$), a variation of $T$ instead of $\mu$ should also affect the ANE \cite{SM}. This effect predicts at low-$T$ an enhancement of the ANE with increasing $T$, which reflects a general property of the Nernst effect, being determined by entropy transport. However, in case of anomalous transport, the increase is expected to continue upon raising $T$ until close to $k_BT\gg \tilde{\mu}$, where the thermal occupation of the Weyl states at energies above and below the Weyl node is practically the same, resulting in a strong reduction of the ANE \cite{Note1}. Remarkably, both the low-$T$ increase and the high-$T$ crossover towards smaller values are observed in our experimental results (Inset of Fig.~\ref{Fig_S_anomalous}). In addition, it is worth to consider that the high-$T$ reduction of the ANE in TaP occurs at higher $T$ with respect to the case of TaAs. This is consistent with band structure calculations, which yield a  $\tilde{\mu}$ about a factor two larger in TaP as compared to TaAs for the lowest lying Weyl point. This property  simulates the influence of the Fermi level with respect to the Weyl points on the anomalous Nernst effect \cite{Note3}.

In conclusion, we reported the observation of a giant ANE in the two Weyl semimetals TaAs and TaP. The occurance of this ANE can be understood as a direct evidence for a large Berry curvature and thus the Weyl points close to the Fermi level, which qualifies Nernst effect measurements as a valuable tool to elucidate Weyl physics. We have shown that the unique identifying feature of the ANE, viz.  the occurance of  field independence beyond a critical field can be traced back to a strong sensitivity of the ANE to field-induced changes of the chemical potential and a Lifshitz transition due to the separation of the Weyl points in momentum space.

\begin{acknowledgments}
F.C. thanks T. Meng for discussion about the theory of Weyl semimetals. We thank M. Richter and C. Timm for valuable discussions. This project has been supported by the Deutsche Forschungsgemeinschaft through SFB1143 (project C07). This project has received funding from the European Research Council (ERC) under the European Unions' Horizon 2020 research and innovation programme (grant agreement No 647276 - MARS - ERC-2014-CoG).
\end{acknowledgments}

\newpage

\end{document}